\documentclass[aps,prl,reprint,superscriptaddress]{revtex4-1}
\usepackage{graphicx}
\usepackage{dcolumn}
\usepackage{bm}
\usepackage{epstopdf}
\usepackage{color}
\usepackage{sidecap}
\usepackage{blindtext}

\begin{document}

\title{Proximity effect induced intriguing superconductivity in van der Waals heterostructure of magnetic topological insulator and conventional superconductor}

 \author{Peng Dong} \thanks{These authors contributed equally to this work.}
\author{Xiang Zhou} \thanks{These authors contributed equally to this work.}
\author{Xiaofei Hou} \thanks{These authors contributed equally to this work.}
\author{Jiadian He}
\author{Yiwen Zhang} 
\author{Yifan Ding}
\author{Xiaohui Zeng}
\author{Jinghui Wang}
\author{Yueshen Wu}
\affiliation{ShanghaiTech Laboratory for Topological Physics \& School of Physical Science and Technology, ShanghaiTech University, Shanghai 201210, China}

\author{Kenji Watanabe}
\affiliation{Research Center for Functional Materials, National Institute for Materials Science, Tsukuba 305-0044, Japan}
\author{Takashi Taniguchi}
\affiliation{International Center for Materials Nanoarchitectonics, National Institute for Materials Science, Tsukuba 305-0044, Japan}

\author{Wei Xia}
\email{xiawei2@shanghaitech.edu.cn}
\author{Yanfeng Guo}
\affiliation{ShanghaiTech Laboratory for Topological Physics \& School of Physical Science and Technology, ShanghaiTech University, Shanghai 201210, China}

\author{Yulin Chen}
\affiliation{ShanghaiTech Laboratory for Topological Physics \& School of Physical Science and Technology, ShanghaiTech University, Shanghai 201210, China}
\affiliation{Department of Physics, Clarendon Laboratory, University of Oxford, Oxford OX1 3PU, UK}

\author{Wei Li}
\email{w\_li@fudan.edu.cn}
\affiliation{State Key Laboratory of Surface Physics and Department of Physics, Fudan University, Shanghai 200433, China}

\author{Jun Li}
\email{lijun3@shanghaitech.edu.cn}
\affiliation{ShanghaiTech Laboratory for Topological Physics \& School of Physical Science and Technology, ShanghaiTech University, Shanghai 201210, China}

\date{\today}

\begin{abstract}
Nontrivial topological superconductivity has received enormous research attentions due to its potential for diverse applications in topological quantum computing. The intrinsic issue concerning the correlation between a topological insulator and a superconductor is, however, still widely open. Here, we systemically report an emergent superconductivity in a cross-junction composed of a magnetic topological insulator MnBi$_2$Te$_4$ and a conventional superconductor NbSe$_2$. Remarkably, the interface indicates existence of a reduced superconductivity at surface of NbSe$_2$ and a proximity-effect-induced superconductivity at surface of MnBi$_2$Te$_4$. Furthermore, the in-plane angular-dependent magnetoresistance measurements reveal the fingerprints of the paring symmetry behaviors for these superconducting gaps as a unconventional nature. Our findings extend our views and ideas of topological superconductivity in the superconducting heterostructures with time-reversal symmetry breaking, offering an exciting opportunity to elucidate the cooperative effects on the surface state of a topological insulator aligning a superconductor.

\end{abstract}

\maketitle

Topological superconductors (TSC), which can host Majorana quasiparticles \cite{Frolov,Sato}, are generally considered as a promising path to obey non-Abelian statistics and encode and manipulate quantum information in a topologically protected manner \cite{Clarke,Zhu}. Commonly, the superconductivity gap symmetry of TSC behaves a nontrivial topological nature and the pair potential of Cooper pairs exhibits an anisotropic orbital symmetry, such as a chiral \emph{p}-wave superconductivity \cite{Ivanov}. Up to now, the most persuasive way to achieve TSC is through the conventional \emph{s}-superconductor proximate to a topological insulator (TI) \cite{Fu,Sun,Lutchyn,Kim,Mourik,Nadj,Jack}. Due to the inevitable degradation of superconducting wavefunction into the TIs in the presence of the interfacial electronic states \cite{Liu2,Geim,Rhodes,Novoselov}, interface-control is challenging for topological superconducting heterojunction engineering.

Recently, the van de Waals magnet MnBi$_2$Te$_4$ (MBT) has attracted great attention owing to the existence of the long-range magnetic order in the nontrivial topological phase \cite{He,Liu,Otrokov,Li,Deng}. MBT crystallizes an alternatively stacking topological insulating Bi$_2$Te$_3$ layers and metallic MnTe layers. As a consequence, the bulk MBT behaves as a Weyl semimetal \cite{Li}. In addition, MBT is an A-type antiferromagnet (AFM) in which the Mn$^{2+}$ ions are ferromagnetically ordered within a single layer (SL) while antiferromagnetically coupled between layers when the temperature is lower than the N$\acute{e}$el temperature. Therefore, of particular interest is that when the topological MBT flake with time-reversal symmetry breaking is proximate to a conventional superconductor (SC), the heterointerface will provide an ideal path to understanding the emergence of rich topological superconducting phenomena. Historically, it has been proposed in theory that a heterojunction comprised of MBT and a conventional $s$-wave SC probably induces a TSC associated with chiral Majorana edge modes. In experiments, the superconducting thin films of Nb and NbN have been grown on MBT \cite{Chen,Xu}, and a proximity-effect-induced superconductivity gap of about 0.1 meV has been observed. More interestingly, a clear Coulomb blockade oscillation has been perceived. Nevertheless, the understanding of the surface state of MBT in proximity to a superconductor remains a complex and challenging area of research, and many questions still lack comprehensive answers.

In this work, we fabricated MBT/NbSe$_2$ heterojunctions by stacking the single crystalline MBT and NbSe$_2$. By selecting different electrodes on the heterojunctions, we simultaneously studied the transport properties of the NbSe$_2$, the MBT, the whole heterojunction, and the interface. Intriguingly, the emergent three superconducting gaps have been perceived through the measurements on the heterojunctions, including the intrinsic NbSe$_2$, the reduced superconductivity at the surface of the NbSe$_2$, and the proximity-effect-induced superconductivity at the surface of the MBT. 

Fig. 1(a) shows the optical microscopic image and the schematic diagram of the MBT/NbSe$_2$ cross-junction, where the superconducting NbSe$_2$flake ($\sim$ 30 nm in thickness) was stacked on top of the magnetic topological MBT flake ($\sim$ 40 nm), configuring a “cross-like” junction, and a 20 nm thick insulating $h$-BN was laid on the bottom of the heterostructure as a substrate. The electrodes on MBT and NbSe$_2$ flakes can be selected in various channels to reveal the electronic properties of the NbSe$_2$, the MBT, the whole heterojunction and its heterointerface. 

Considering the interface between the MBT and NbSe$_2$, as given in Fig. 1(b), MBT reveals a ferromagnetic order along the out-of-plane, while electrons in NbSe$_2$ are condensed into Cooper pairs which behave as a singlet superconducting state ($\uparrow \downarrow$-$\downarrow  \uparrow$) below the critical temperature ($T_c$). As has been demonstrated before\cite{buzdin2005,Xi,Hess,Hamill,Yokoya}, when the distance from NbSe$_2$ increases, its superconducting gap $\Delta_0$ will gradually degenerate and form a reduced superconductivity gap $\Delta_1$ at the interface, as illustrated in Fig. 1(c). Notably, the gap $\Delta_1$ strongly depends on the electron state of the non-SC side. Once the non-SC side is a normal metal, the superconducting wave function from the NbSe$_2$ can flow into the metal within a distance less than the coherence length $\xi_{\text{N}}=\sqrt{\hbar D/k_{\text{B}}T}$, where $D$ is the diffusion coefficient \cite{buzdin2005}. While for a magnetic material, the superconductivity coherence length ($\xi_{\text{M}}$) turns into $\xi_{\text{M}}= \sqrt{\hbar D_{\text{M}}/k_{\text{B}}T_{\text{M}}}$, where $T_{\text{M}}$ is the magnetic transition temperature. Meanwhile, the superconducting wave function decays exponentially $\Delta_{\text{M}}$=exp$(D_{\text{M}}⁄\xi_{\text{M}})$ due to the existence of the exchange splitting from the spin-spin interaction, resulting in a restricted superconducting gap function ($\Delta_2$). Since the exchange splitting is absent in an AFM, the superconductivity coherence length should be comparable to that of normal metal. However, due to the A-type AFM magnetic structure of MBT, it is essential to take into account the ferromagnetic coupling when considering the interface between MBT and NbSe$_2$.

\begin{figure}[!htbp]
\includegraphics[width=1\linewidth,clip]{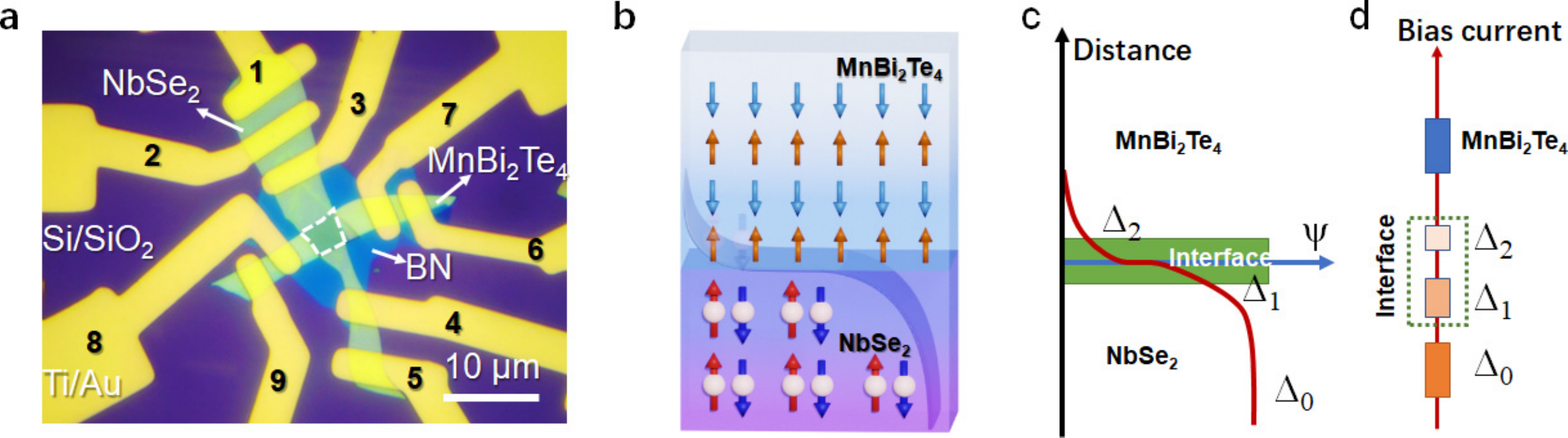}
\caption{(color online).  Sample geometry and the interface of MBT/NbSe$_2$. (a) The optical microscopic image of the MBT/NbSe$_2$ structure. The bottom h-BN (~20 nm) provides an atomic flat substrate, and MBT (40 nm) and NbSe$_2$ (30 nm) flakes are stacked as a heterojunction, in which the junction area is labeled by a white dash line. Nine electrodes marked from 1 to 9 are connected in different configurations of current flowing and voltage measurement for comparison of electrical properties of the pristine NbSe$_2$ (current 1-5, voltage 2-3), MBT (6-9, 7-8), the whole heterojunction (1-6, 3-8), and the interface (1-6, 5-9) channels. (b) Schematic images of MBT/NbSe$_2$ heterojunction. MBT reveals an antiferromagnetic order along the out-of-plane, while electrons in NbSe$_2$ demonstrate Cooper pairs with opposite spin directions in the superconducting state. (c) Schematic image of the degradation of the superconducting wave function ($\psi$) along the distance from NbSe$_2$. Here, $\Delta_0$, $\Delta_1$, and and $\Delta_2$ correspond to the superconductivity gaps of the intrinsic NbSe$_2$, the surface superconductivity of NbSe$_2$, and the proximity-effect induced superconductivity at the MBT surface. (d) The equivalent circuit of the junction.
}
\end{figure}

By selecting electrodes on the MBT and NbSe$_2$ flakes in different channels (see Fig. 1(a)), we were able to evaluate the electronic properties of NbSe$_2$, MBT, the whole heterojunction, and the interface are illustrated in Fig. 2(a), 2(e), 2(i), and 2(m). The detailed analysis is introduced in the Supplementary Materials. 

For the pristine NbSe$_2$, the temperature- and magnetic field-dependent resistances are shown in Fig. 2(b) and 2(c), respectively, indicating the intrinsic superconductivity of NbSe$_2$. Specifically, a superconducting transition temperature ($T_{\text{c}}$) of 7 K and a critical field ($B_{\text{c}}$) of 3 T at 1.8 K were observed. Moreover, the critical current of NbSe$_2$ is about 1.8 mA at 1.8 K, as shown in Fig. 2(d). According to the electron-phonon coupling strength in BCS model  $2 \Delta_0=3.52k_{\text{B}}T_{\text{c}}$, we can estimate the superconducting gap of the pristine NbSe$_2$ to be 1.2 meV at $T$ = 0, which is consistent with the previous reports\cite{Hess}. 

The transport properties of MBT are consistent with previous results\cite{Otrokov,Zeugner,Cui,Lee}, where the N$\acute{e}$el temperature ($T_{\text{N}}$) is 23.7 K (see Fig. 2(f)) and the spin-flop critical field ($B_{\text{c}}^{\text{MBT}}$) is $\sim$3.34 T (see Fig. 2(g)). It seems that the $B_{\text{c}}^{\text{MBT}}$ is comparable to that of the $B_{\text{c}}$ of NbSe$_2$ ($\sim$3 T). While, one can distinguish the contribution of MBT or NbSe$_2$ because these transitions induce opposite modifications on the magnetoresistance, namely, MBT causes a downturn while NbSe$_2$ leads to an upturn as the magnetic field is increased (see Fig. 2(c) and Fig. 2(g), respectively).

\begin{figure*}[!htbp]
\includegraphics[width=0.7\textwidth,clip]{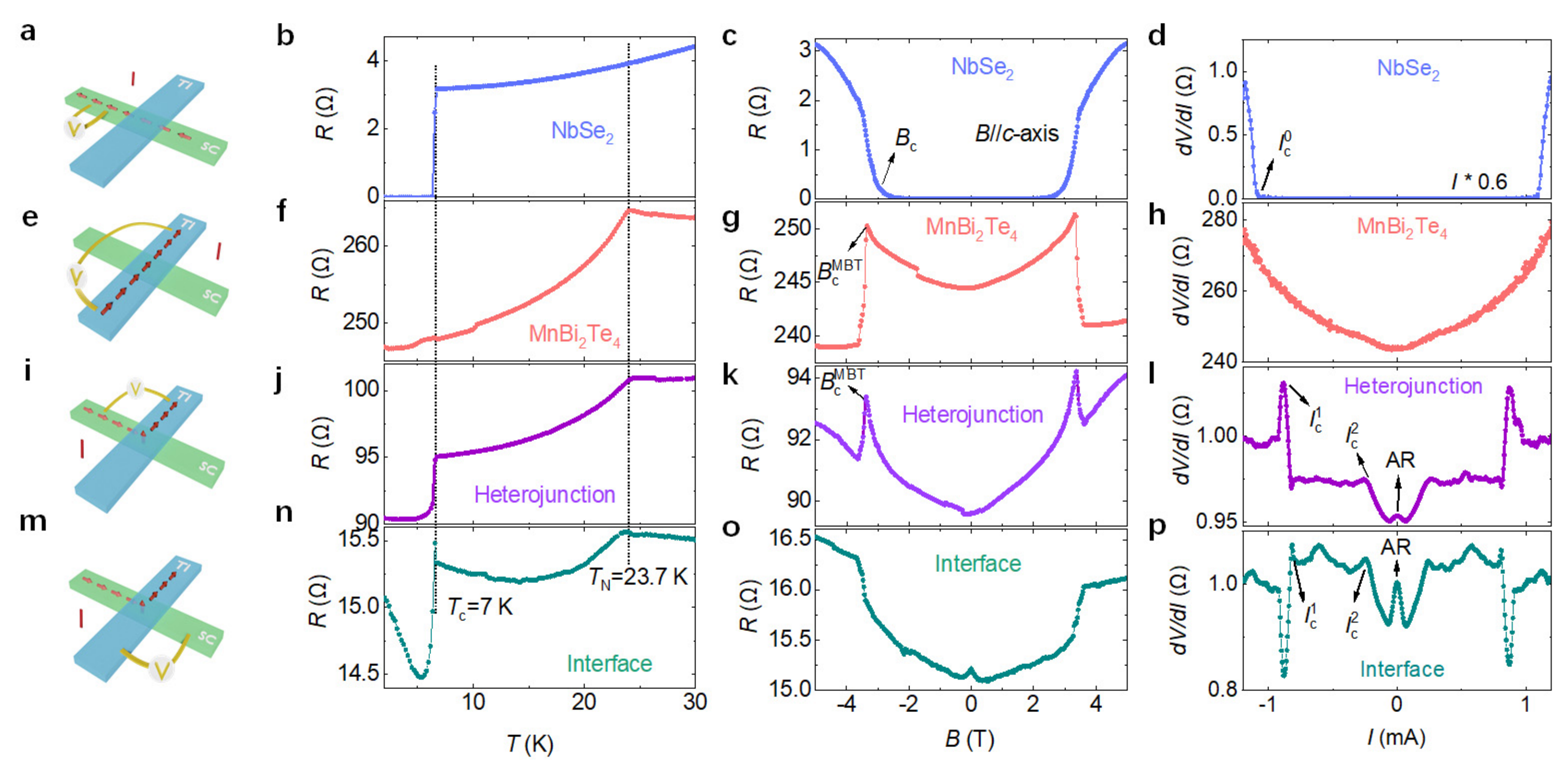}
\caption{(color online). Transport properties for different channels of the heterojunction. (a), (e), (i), and (m). Measurement schematic of the NbSe$_2$ (current 1-5, voltage 2-3 as shown in Fig. 1), MBT (6-9, 7-8), the whole heterojunction (1-6, 3-8), and the interface (1-6, 5-9) channels. (b), (f), (j), and (n). Temperature-dependent resistance for NbSe$_2$, MBT, MBT/NbSe$_2$ heterojunction, and interface, respectively. Here, all $R-T$ curves are measured under zero field. Particularly, two temperature transitions are identified at the R-T curve of the heterojunction and interface, including the superconducting transition Tc of NbSe$_2$ at 7 K and antiferromagnetic transition $T_{\text{N}}$ of MBT at 23.7 K. The corresponding magnetoresistances at 1.8 K are given in (c), (g), (k), and (o), where the field was applied along the out-of-plane. The magnetic transition of MBT is observed in (g) and (k) as the spin-flop critical field ($B_{\text{c}}^{\text{MBT}}$), and the critical field of NbSe$_2$ exhibits in (c) as $B_{\text{N}}$. (d), (h), (l), and (p). The $dV/dI$ vs. $I$ curves at 1.8 K for NbSe$_2$, MBT, MBT/NbSe$_2$ heterojunction, and the interface, respectively. Here, the current axis for NbSe$_2$ in (d) is timed 0.6, because the critical current of NbSe$_2$ is 1.8 mA.
}
\end{figure*}

With the whole heterojunction measurement configuration (see Fig. 2(i)), the temperature-dependent resistance shows two characteristic transitions at 23.7 and 7 K in Fig. 2(j), which correspond to the $T_{\text{N}}$ of MBT and the $T_{\text{c}}$ of NbSe$_2$, respectively. The magnetoresistance of the heterojunction in Fig. 2(k) reveals a downturn at 3.34 T following a nonlinear upturn transition at around 3 T, which can be related to the critical field of MBT ($B_{\text{c}}^{\text{MBT}}$) and NbSe$_2$  ($B_{\text{c}}$), respectively. On the other hand, the $dV/dI$ vs. $I$ curve of the heterojunction at 1.8 K (see Fig. 2(i)) demonstrates two pairs of transitions, locating at around $\pm$895 and $\pm$ 245 $\mu$A, respectively, which will be discussed in detail in the following part.

The transport properties of the MBT/NbSe$_2$ interface were investigated by applying a current through two adjacent sides of the “cross-shape” heterojunction and measuring the voltage through the other two sides (see Fig. 2(m)). The temperature dependence of resistance in Fig. 2(n) also shows two distinct transitions at 23.7 and 7 K, corresponding to the $T_{\text{N}}$ and $T_{\text{c}}$, respectively. However, only an upturn transition is observed on the magnetoresistance curve as shown in Fig. 2(o), suggesting that the superconducting transition dominates the magnetoresistance at the interface. It is worth noting that the resistance of the interface exhibits an upturn behavior below the superconductivity transition, which can be attributed to the Andreev reflection (AR) effect. Such AR phenomenon also contributes to the peak of differential resistance at the zero-bias current as shown in Fig. 2(p). On the other hand, two pairs of transitions are observed at  $\pm$878 and $\pm 248 \mu$A on the $dV/dI$ - $I$ curve, which are consistent with those of heterojunction. However, these critical currents are significantly lower than that of the pristine NbSe$_2$, which is 1.8 mA at 1.8 K. These features can be attributed to the superconducting transitions of the reduced superconductivity at the surface of NbSe$_2$ and the proximity-effect-induced areas on the surface of MBT (see Fig. 1(d)).

\begin{figure*}[!htbp]
\includegraphics[width=0.7\textwidth,clip]{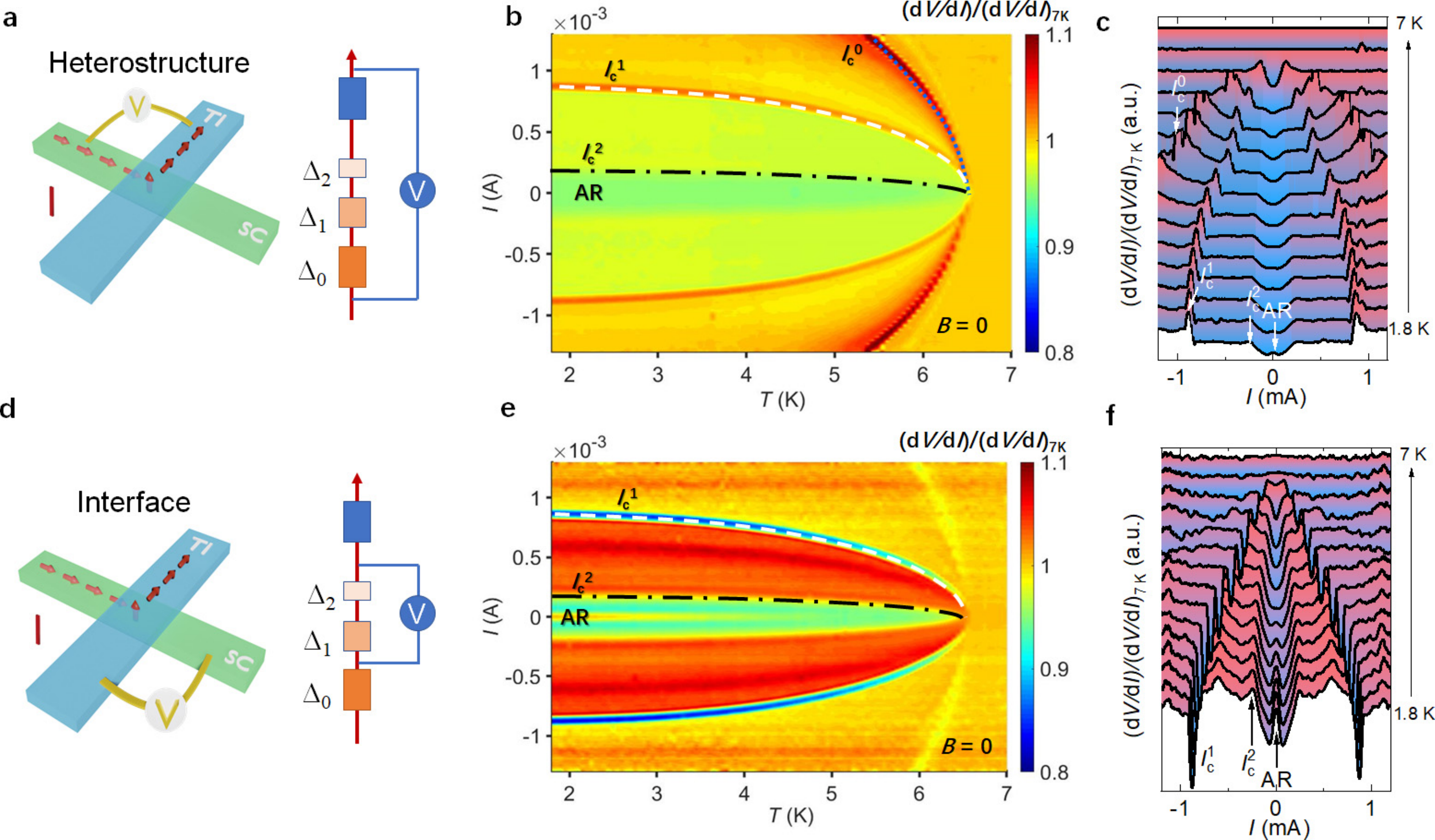}
\caption{(color online). Temperature-dependent critical current. Measurement schematic of (a) the whole MBT/NbSe$_2$ heterojunction and (d) the MBT/NbSe$_2$ interface. The differential resistance mapping of (b) the heterojunction and (e) the interface with the temperature ranging from 1.8 K to 7.0 K. The selected dV/dI-I spectrums of (c) the heterojunction and (f) the interface with several typical temperatures. $I_{\text{c}}^0$, $I_{\text{c}}^1$, and $I_{\text{c}}^2$ correspond to the critical currents of the intrinsic NbSe$_2$, the reduced superconductivity at the surface of the NbSe$_2$, and the proximity-effect-induced superconductivity at the surface of the MBT.
}
\end{figure*}

\begin{figure*}[!htbp]
\includegraphics[width=0.7\textwidth,clip]{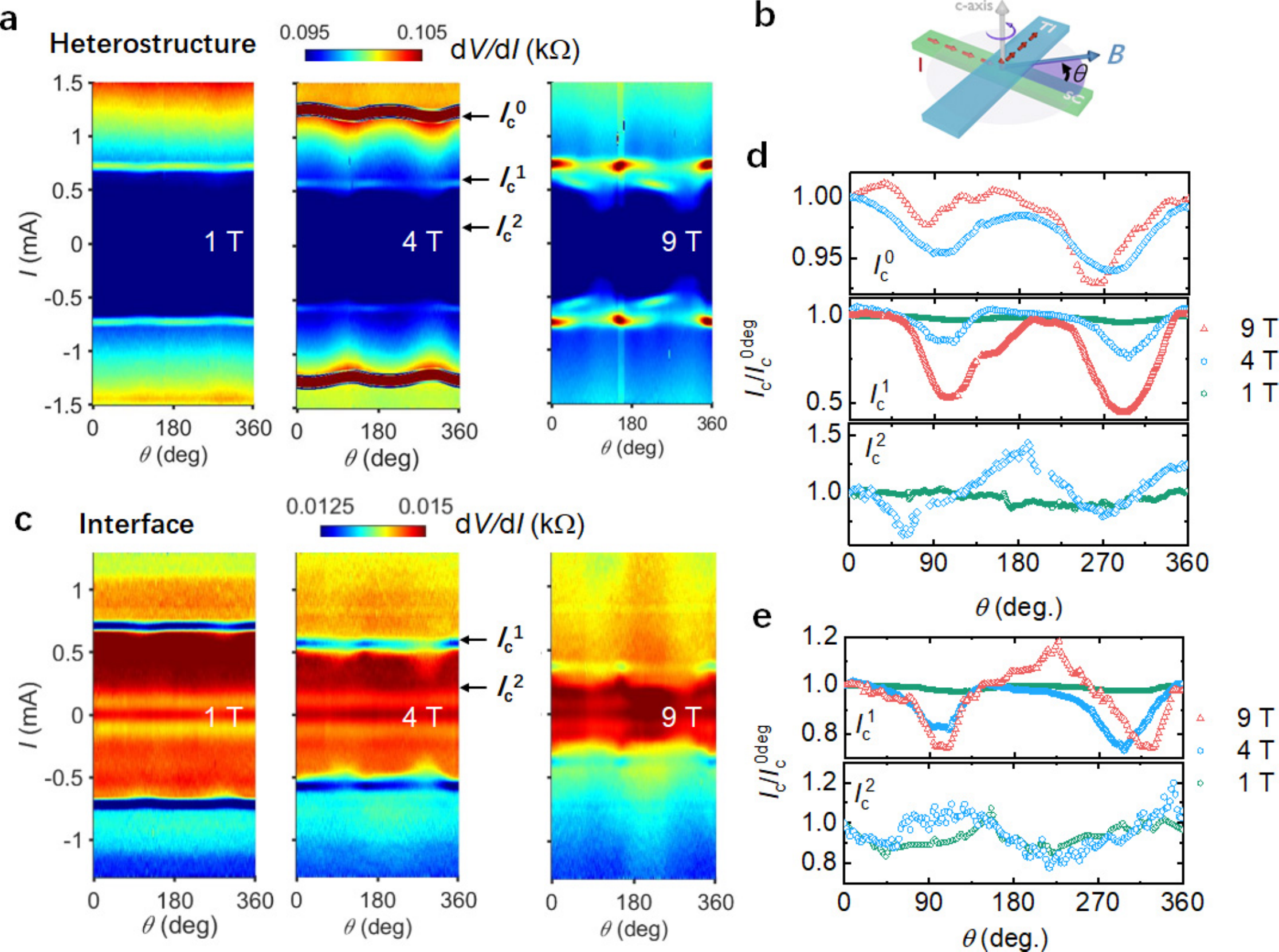}
\caption{(color online). Angular-dependent critical current at 1.8 K. (b) Schematic image of the measurement setup. Here, the magnetic field was applied within the in-plane, and $\theta$ corresponds to the angle between the current within the NbSe$_2$ and field $B$. (a) and (c). The angular-dependent differential resistance mapping under different magnetic fields for the whole heterojunction and the interface, respectively. (d) and (e). The angular-dependent of $I_{\text{c}}^0$, $I_{\text{c}}^1$, and $I_{\text{c}}^2$  at 4 T for the whole heterojunction and the interface, respectively. Here, the $I_{\text{c}}^0$, $I_{\text{c}}^1$, and $I_{\text{c}}^2$ are estimated from mapping in (a) and (b). 
}
\end{figure*}

To understand the proximity-effect-induced superconductivity and AR through the heterojunction, we further studied the temperature-dependent $dV/dI-I$ spectrums of the whole heterojunction and the interface. Fig. 3(b) shows the mapping of normalized differential resistance at various temperatures of the whole heterostructure, and the differential resistances at some typical temperatures are given in Fig. 3(c). Three characteristic contour lines, labeled as $I_{\text{c}}^0$, $I_{\text{c}}^1$, and $I_{\text{c}}^2$, can be seen on both positive and negative current sides of the mapping in Fig. 3(c), corresponding to the peaks of the $dV/dI$-$I$ curves shown in Fig. 3(d). It is worthy to mention that the temperature-dependent $I_{\text{c}}^0$, $I_{\text{c}}^1$, and $I_{\text{c}}^2$ and their corresponding gaps ($\Delta_{\text{c}}^0$, $\Delta_{\text{c}}^1$, and $\Delta_{\text{c}}^2$, respectively) are consistent with the conventional BCS model as $\Delta_{\text{T}}=1.74 \Delta_{\text{c}}^0 \sqrt{1-T/T_{\text{c}}}$. Therefore, $I_{\text{c}}^0$, $I_{\text{c}}^1$, and $I_{\text{c}}^2$, can be attributed to the intrinsic NbSe$_2$, the reduced superconductivity at the surface of the NbSe$_2$, and the proximity-effect-induced superconductivity at the surface of the MBT. 

In comparison, we perform the same measurement on the interface (see Fig. 3(d)). The mapping results shown in Fig. 3(e) reveal the absence of  $I_{\text{c}}^0$, which further confirms that $I_{\text{c}}^0$ corresponds to the intrinsic NbSe$_2$ since it is excluded during the measurement setup. Moreover, the transition of $I_{\text{c}}^1$ and $I_{\text{c}}^2$, and particularly the AR, are considerably more intense than those of the heterojunction. The main reason is that both NbSe$_2$ and MBT sides are excluded in this configuration, and the interfacial effect dominates the measurement result. Therefore, the “cross-junction” geometry is an ideal configuration to investigate the interfacial effects.

Considering the AR effect at an SC/non-SC interface, it is an effective method to determine the polarization $P$ of a certain material \cite{Soulen,Strijkers}. In our present AR effect in the MBT/NbSe$_2$ interface, the reduction of conductance at zero-bias current indicates weak spin-polarization from MBT. However, the MBT crystal in our case is about 40 nm ($\sim$10 SLs), which should demonstrate a long-term AFM coupling as a bulk crystal. For comparison, we also studied the AR behavior of the Fe$_5$GeTe$_2$/NbSe$_2$, where Fe$_5$GeTe$_2$ reveals typical FM coupling along the $c$-axis. Surprisingly, the $dI/dV$-$V$ spectrums at the interface of MBT/NbSe$_2$ and Fe$_5$GeTe$_2$/NbSe$_2$ (see Fig. S5) are comparable. Therefore, we can conclude that the surface state of MBT reveals an FM ordering and partially restricts AR. 

According to the temperature- and field-dependent (see Fig. 2 and Figs. S1-S3) results, we have confirmed the observation of proximity-effect-induced superconductivity ($\Delta_1$) on the surface of MBT. In principle, the AFM coupling and topological nontrivial state in the electron bands structure may contribute to the superconducting coupling, resulting in a novel superconductivity gap symmetry or even topological superconductivity, which should be pronouncedly different from the conventional superconductivity in NbSe$_2$. To uncover the symmetry of the proximity-effect-induced superconductivity in MBT, we also measured the angular-dependent magnetoresistance spectrums of the whole MBT/NbSe$_2$ heterojunction and the interface. 

The angular-dependent measurement geometry is shown in Fig. 4(b). Figs. 4(a) and 4(c) show the mapping of angular-dependent $dV/dI$-$I$ spectrums for the heterostructure and interface at 1.8 K under magnetic fields of 1, 4, and 9 T. In particular, the  $I_{\text{c}}^0$, $I_{\text{c}}^1$, and $I_{\text{c}}^2$ can be identified. Moreover, the corresponding angular-dependent normalized critical currents are plotted in Figs. 4(d) and 4(e), where $\theta$ is the angle between the magnetic field and $a$-axis direction of NbSe$_2$. Surprisingly, in contrast to what is generally considered as a conventional $s$-wave superconductor that should behave as an isotropic superconductivity symmetry, the angular-dependent $I_{\text{c}}^0$ of NbSe$_2$  reveals a cosine function with two-fold symmetry, where the maximum appears at 0$^{\circ}$ and 180$^{\circ}$ and the minimum appears at 90$^{\circ}$ and 270$^{\circ}$, respectively. Thus, an unconventional pairing mechanism may exist in finite layers of NbSe$_2$, where spontaneous nematic superconductivity or strong gap-mixing triggered by a small symmetry-breaking field may induce an anisotropic nature \cite{Huang}. However, a deeper understanding of the symmetry behavior is still needed. 

The angular dependence of $I_{\text{c}}^1$ is comparable for both heterostructure and interface measurement configurations (see Figs. 4(d) and 4(e)). Nevertheless, it does not appear to follow a cosine or sine function, but rather a combination of trigonometric function with a four-fold symmetry. Although, the minimum appears at about 90$^{\circ}$ and 270$^{\circ}$ as that of NbSe$_2$, the maximum appears at 120$^{\circ}$ and 330$^{\circ}$ ($B$ = 4 T), indicating that the phase of $I_{\text{c}}^1$ is shifted compared to  $I_{\text{c}}^0$. Moreover, the $I_{\text{c}}^2 - \theta$ curves also reveal a two-fold symmetry for both measurement configurations, while the pattern is slightly different from those of $I_{\text{c}}^0$ and $I_{\text{c}}^1$. Particularly, the minimum of $I_{\text{c}}^2$ appears at $\sim$45$^{\circ}$ and 270$^{\circ}$, which reveals a 45$^{\circ}$ phase shift from both $I_{\text{c}}^0$ and $I_{\text{c}}^1$. However, the maximum seems to correspond to that of $I_{\text{c}}^0$ and $I_{\text{c}}^1$. Based on the angular-dependent profiles of $I_{\text{c}}^0$, $I_{\text{c}}^1$, and $I_{\text{c}}^2$, although we can hardly confirm the exact gap symmetry of the induced superconductivity on the surface of MnBi$_2$Te$_4$ or even the intrinsic NbSe$_2$, the superconductivity gap symmetries of $\Delta_1$ and $\Delta_2$ are basically different from that of NbSe$_2$ ($\Delta_0$).

In summary, we studied the van der Waals “cross-shape” heterojunction between a nontrivial TI MBT and a conventional SC NbSe$_2$. Through different measurement configurations, three different superconductivity gaps have been observed, including the intrinsic NbSe$_2$, the reduced superconductivity at the surface of the NbSe$_2$, and the proximity-effect-induced superconductivity at the surface of the MBT. A clear zero-bias peak was observed in the $dV/dI-I$ curve, which is corresponding to the AR effect at a FM/SC interface, suggesting that the surface of MBT exhibits a ferromagnetic plane. The temperature-, field- and angular-dependent $dV/dI-I$ spectrums with the whole heterojunction and the interface measurement configurations have been systematically investigated. Although we cannot identify the superconductivity pairing symmetry, an unconventional superconducting state which is different from the conventional SC NbSe$_2$ should be seriously considered, as the various quantum states like topological states and magnetic ordering may contribute to the electronic coupling. 

This research was supported in part by the Ministry of Science and Technology (MOST) of China (No. 2022YFA1603903), the National Natural Science Foundation of China (Grants No. 12004251, 12104302, 12104303, 12304217), the Science and Technology Commission of Shanghai Municipality, the Shanghai Sailing Program (Grant No. 21YF1429200), the start-up funding from ShanghaiTech University, and Beijing National Laboratory for Condensed Matter Physics, the Interdisciplinary Program of Wuhan National High Magnetic Field Center (WHMFC202124), the open project from State Key Laboratory of Surface Physics and Department of Physics, Fudan University (Grant No. KF2022-13), the Shanghai Sailing Program (23YF1426900), and the Double First-Class Initiative Fund of ShanghaiTech University. Growth of $h$-BN was supported by the Elemental Strategy Initiative conducted by the MEXT, Japan (JPMXP0112101001), JSPS KAKENHI (JP20H00354) and A3 Foresight by JSPS.


%

\end{document}